\documentclass[envcountsame]{llncs}
\usepackage{amssymb}
\usepackage{amsmath}
\usepackage{graphicx}
\usepackage{color}
\usepackage{algorithm, algorithmic}
\usepackage{tikz}
\usepackage{colortbl}
\usepackage{hyperref}

\usetikzlibrary{arrows,shapes,chains,plotmarks,positioning,calc}

\def\lz#1{\textbf{\color{blue} <<< LZ: #1 >>>}}

\newcommand{\Xs}{\mathbf{X}_n}
\newcommand{\Real}{\mathbb{R}}
\newcommand{\Xstuple}{X_1, \ldots, X_n}
\newcommand{\xs}{\mathbf{x}_n}

\newcommand{\zs}{\mathbf{z}_n}
\newcommand{\zstuple}{z_1, \ldots, z_n}
\newcommand{\while}{\textsf{while}}
\newcommand{\progskip}{\textsf{skip}}
\newcommand{\abort}{\textsf{abort}}
\newcommand{\progif}{\textsf{if}}
\newcommand{\progelse}{\textsf{else}}
\newcommand{\progthen}{\textsf{then}}

\begin{document}
	
	\title{Finding Polynomial Loop Invariants \\ for Probabilistic Programs} 
	\author{
		Yijun Feng\inst{1}, Lijun Zhang\inst{2}, David N. Jansen\inst{2} (0000-0002-6636-3301), Naijun Zhan\inst{2}, and Bican Xia\inst{1}
		\institute{
			LMAM \& School of Mathematical Sciences, Peking University, Beijing, China \and
			State Key Laboratory of Computer Science,
			Institute of Software, CAS, Beijing, China}
}
	
\maketitle
	
\begin{abstract}
  Quantitative loop invariants are an essential element in the verification of
  probabilistic programs. Recently, multivariate Lagrange
  interpolation has been applied to synthesizing polynomial
  invariants.
  In this paper, we propose an alternative approach.
  First, we fix a polynomial template as a candidate of a loop invariant.
  Using Stengle's Positivstellensatz and a transformation to a sum-of-squares problem,
  we find sufficient conditions on the coefficients.
  Then, we solve a semidefinite programming feasibility problem
  to synthesize the loop invariants.
  If the semidefinite program is unfeasible,
  we backtrack after increasing the degree of the template.
  Our approach is semi-complete in the sense
  that it will always lead us to a feasible solution if one exists and numerical errors are small.
  Experimental results show the efficiency of our approach.
\end{abstract}

\section{Introduction}
Probabilistic programs extend standard programs with probabilistic choices
and are widely used in protocols, randomized algorithms, stochastic games, etc.
In such situations, the program may report incorrect results with a certain probability,
rendering classical program specification methods~\cite{Dijkstra3,hoare8} inadequate.
As a result, formal reasoning about the correctness needs to be based on quantitative specifications.
Typically, a probabilistic program consists of steps that choose probabilistically between several states.
and the specification of a probabilistic program contains constraints on the probability distribution of final states,
e.\,g.\@ through the expected value of a random variable.
Therefore the expected value is often the object of correctness verification~\cite{McIver-book,katoen2010linear9,Henzinger}.

To reason about correctness for probabilistic programs, quantitative annotations are needed. Most importantly, correctness of while loops can be proved by inferring special bounds on expectations,
usually called \emph{quantitative loop invariants}~\cite{McIver-book}.
As in the classical setting, finding such invariants is the bottleneck of proving program correctness.
For some restricted classes, such as linear loop invariants,
some techniques have been established~\cite{morgan1996probabilistic,katoen2010linear9,chakarov1}.
To use them to synthesize polynomial loop invariants,
so-called linearization can be used~\cite{barthe2016synthesizing},
a technique widely applied in linear algebra.
It views higher-degree
monomials as new variables, establishes their relationship with
existing variables, and then exploits linear loop invariant generation
techniques. However, the number of monomials grows exponentially when the
degree increases.
Kapur et al.~\cite{rodriguez2007generating} introduce solvable mappings,
which are a generalization of affine mappings,
to avoid non-polynomial effects generated by polynomial programs.
Recently, Chen et al.~\cite{chen2015counterexample} applied multivariate Lagrange interpolation
to synthesize polynomial loop invariants directly.

Another important problem for probabilistic programs is the \emph{almost-sure termination problem,}
answering whether the program terminates almost surely.
In~\cite{ferrer2015probabilistic}, Fioriti and Hermanns argued that Lyapunov ranking functions,
used in non-probabilistic termination analysis,
cannot be extended to probabilistic programs.
Instead, they extended the ranking
supermartingale approach~\cite{chakarov2013probabilistic} to the
bounded probabilistic case, and provided a compositional and sound
proof method for the almost-sure termination problem.
In~\cite{kaminski2015hardness}, Kaminski and Katoen
investigated the computational complexity of computing expected outcomes
(including lower bounds, upper bounds and exact expected outcomes) and
of deciding almost-sure termination of probabilistic programs.
In~\cite{ChatterjeePOPL16}, Chatterjee et al.\@ further investigated termination problems for affine probabilistic programs.
Recently, they also presented a method~\cite{chatterjee1604}
to efficiently synthesize ranking supermartingales through Putinar's Positivstellensatz~\cite{putinar1993positive}
and used it to prove the termination of probabilistic programs.
Their method is sound and semi-complete over a large class of programs.

In this paper, we develop a technique exploiting \emph{semidefinite programming} through another Positivstellensatz
to synthesize the quantitative loop invariants.
Positivstellens\"atze are essential theorems in real algebra
to describe the structure of polynomials that are positive (or non-negative) on a \emph{semialgebraic set.}
While our approach shares some similarities with the one in~\cite{chatterjee1604},
the difference to the termination problem requires a variation of the theorem.
In detail, Putinar's Positivstellensatz deals with the situation when the polynomial is strictly positive on a quadratic module,
which is not enough for quantitative loop invariants.
In the program correctness problem, equality constraints are taken into consideration as well as inequalities.
Therefore in our method, Stengle's Positivstellensatz~\cite{stengle1974nullstellensatz} dealing with general real semi-algebraic sets is being used.

As previous results~\cite{katoen2010linear9,gretz2013prinsys6,chen2015counterexample},
our approach is \emph{constraint-based}~\cite{Colon-Cav}.
We fix a polynomial template for the invariants with a fixed degree and generate constraints from the program.
The constraints can be transformed into an emptiness problem of a semialgebraic set.
By Stengle's Positivstellensatz~\cite{stengle1974nullstellensatz}, it suffices to solve a semidefinite programming feasibility problem,
for which efficient solvers exist.
From a feasible solution
(which may not be the tightest one)
we can then obtain the corresponding coefficients of the template.
We verify the correctness of the template.
If the solver does not provide a feasible solution or if the coefficients are not correct,
we refine the analysis by adding constraints to block the undesired solutions 
and get a tighter invariant
or increasing the degree of the template,
which will always lead us to a feasible solution if one exists.

The method is applied to several case studies taken from~\cite{chen2015counterexample}.
The technique usually solves the problem within one second,
which is about one tenth of the time taken by the tool described in~\cite{chen2015counterexample}.
Our tool supports real variables rather than discrete ones,
and supports programs that require polynomial invariants.
We il\-lu\-strate these features by analyzing a non-linear perceptron program and a model for airplane delay with continuous distributions.
Moreover, we conduct a sequence of trials on parameterized probabilistic programs
to show that the main influence factor on the running time of our method is the degree of the invariant template.
We compare our results on these examples with the Lagrange Interpolation Prototype (LIP) in~\cite{chen2015counterexample},
\textsc{Prinsys}~\cite{gretz2013prinsys6}
and the tool for super-martingales (TM)~\cite{chakarov2013probabilistic}.


\section{Preliminaries}\label{sec:pre}

In this section we introduce some notations.
We use $\Xs$ to denote an $n$-tuple of variables $(\Xstuple)$.
For a vector ${\alpha} = (\alpha_1, \ldots, \alpha_n) \in \mathbb{N}^n$,
$\Xs^{\alpha}$ denotes the monomial $X_1^{\alpha_1}\cdots X_n^{\alpha_n}$,
and $d = \sum_i \alpha_i$ is its degree.
\begin{definition}
	A polynomial $f$ in variables $\Xstuple$ is a finite linear combination of monomials:
	$
	f=\sum_{\alpha}c_{\alpha}\Xs^{\alpha}
	$
	where finitely many $c_{\alpha} \in \Real$ are non-zero.
\end{definition}
The degree of a polynomial is the highest degree of its component monomials.
Extending the notation, for a sequence of polynomials
$\mathcal{F} = (f_1,\ldots,f_s)$ and a vector $\alpha=(\alpha_1, \ldots, \alpha_s) \in \mathbb{R}^s$,
we let $\mathcal{F}^{\alpha}$ denote $\prod_{i=1}^s f_i^{\alpha_i}$.
The polynomial ring with $n$ variables is denoted with $\Real[\Xs]$,
and the set of polynomials of degree at most $d$ is denoted with $\Real^{\leq d}[\Xs]$.
For $f \in \Real[\Xs]$ and $\zs=(\zstuple)\in\Real^n$, $f(\zs)\in\Real$ is the value of $f$ at $\zs$.

A \textit{constraint} is a quantifier-free formula constructed from (in)equalities of polynomials.
It is \textit{linear} if it contains only linear expressions.
A semialgebraic set is a set described by a constraint:
\begin{definition}\label{def:semialgebraset}
	A semialgebraic set in $\Real^k$ is a finite union of sets of the form
	$\{x\in \Real^k| f(x)=0 \mathrel{\wedge} \bigwedge_{g\in\mathcal{G}} g>0\}$,
	where $f$ is a polynomial and $\mathcal{G}$ is a finite set of polynomials.
\end{definition}

A polynomial $p(\Xs)\in \Real[\Xs]$ is a \textit{sum of squares} (or SOS, for short),
if there exist polynomials $f_1(x),\ldots ,f_m(x)\in \Real[\Xs]$ such that
$
p(\Xs)=\sum_{i=1}^m f_i^2(\Xs)
$.
Chapters~2 and 3 of~\cite{blekherman2012semidefinite} introduce a way to transform the problem whether a given polynomial is an SOS
into a \textit{semidefinite programming problem} (or SDP, for short),
which is a generalization of linear programming problem.
We introduce the transformation and SDP problems briefly in Appendices~\ref{subsec:sos} and \ref{app:sdp}.

\subsection{Probabilistic Programs}\label{subsec:pp}
We use a simple \emph{probabilistic guarded-command language}
to construct \textit{probabilistic programs} with the grammar:
\[
P::= \progskip\mid\abort\mid x:=E\mid P;P\mid P\ [p]\ P\mid \progif\ (G) \ \progthen \ \{P\}\  \progelse\  \{P\}\mid \while(G)\{P\}
\]
where $G$ is a Boolean expression and $E$ is a real-valued expression defined by the grammar:
\[
\begin{array}{rll}
E&::=c\mid \xs \mid r\mid &\qquad\textrm{constant/variable/random variable}\\
&\phantom{::=} E + E\mid E \cdot E\mid &\qquad\textrm{arithmetic}\\
G&::= E < E\mid G \mathrel{\wedge} G\mid \neg G \quad &\qquad\textrm{guards}
\end{array}
\]
 Random variable $r$ follows a given probability distribution, discrete or continuous. For $p \in [0,1]$, the \textit{probabilistic choice command} $P_0\ [p]\ P_1$ executes $P_0$ with probability $p$ and $P_1$ with probability $1-p$.

\begin{example}
	The following probabilistic program $P$ describes a simple game:
	$$z:=0;\ \while(0<x<y)\ \{x:=x+1[0.5]x:=x-1;\ z:=z+1\}.
	$$
	The program models a game where a player has $x$ dollars at
        the beginning and keeps tossing a coin with probability
        $0.5$. The player wins one dollar if he tosses a head and loses
        one dollar for a tail. The game ends when the player loses all
        his money, or he wins $y-x$ dollars for a predetermined
        $y$. The variable $z$ records the number of tosses made by the
        player during this game.
\end{example}

We assume that the reader is familiar with the basic concepts of probability theory and in particular expectations,
see e.\,g.~\cite{feller1968introduction} for details.
Expectations are typically functions from program states (i.\,e.\@ the real-valued program variables) to $\mathbb{R}$.
An expectation is called a \textit{post-expectation} when it is to be evaluated on the final distribution,
and it is called a \textit{pre-expectation} when it is to be evaluated on the initial distribution.
Let $\mathit{preE}$, $\mathit{postE}$ be expectations and $\mathit{prog}$ a probabilistic program.
We say that the sentence $\langle \mathit{preE} \rangle\ \mathit{prog}\ \langle \mathit{postE} \rangle$ \textit{holds}
if the expected value of $\mathit{postE}$ after executing $prog$ is equal to or greater than the expected value of $\mathit{preE}$.
When $\mathit{postE}$ and $\mathit{preE}$ are functions, the comparison is executed pointwise.

Classical programs can be viewed as special probabilistic programs in the following sense.
For classical precondition $\mathit{pre}$ and postcondition $\mathit{post}$,
let the characteristic function $\chi_\mathit{pre}$ equal 1 if the precondition is true and 0 otherwise,
and define $\chi_\mathit{post}$ similarly.
If one considers a Hoare triple $\{\mathit{pre}\}\ \mathit{prog}\ \{\mathit{post}\}$
where $\mathit{prog}$ is a classical program,
then it holds if and only if $\langle\chi_\mathit{pre}\rangle\ \mathit{prog}\ \langle\chi_\mathit{post}\rangle$ holds in the probabilistic sense.

\subsection{Probabilistic Predicate Transformers}
Let $P_0$, $P_1$ be probabilistic programs, $E$ an expression, $\mathit{post}$ a post-expectation, $\mathit{pre}$ a pre-expectation, $G$ a Boolean expression, and $p\in (0,1)$.
The probabilistic predicate transformer $\mathit{wp}$ can be defined as follows~\cite{gretz2014operational}:
\[
\begin{array}{l}
\mathit{wp}(\progskip, \mathit{post})=\mathit{post}\\
\mathit{wp}(\abort, \mathit{post})=0\\
\mathit{wp}(x:=E, \mathit{post})=\mathit{post}[x/E_{S}(E)]\\
\mathit{wp}(P;\ Q,\mathit{post})=\mathit{wp}(P,\mathit{wp}(Q,\mathit{post}))\\
\mathit{wp}(\progif(G)\ \progthen (P)\ \progelse(Q),\mathit{post})=\chi_G \cdot \mathit{wp}(P,\mathit{post})+(1-\chi_G)\cdot \mathit{wp}(Q,\mathit{post})\\
\mathit{wp}(P\ [p]\ Q,\mathit{post})=p\cdot \mathit{wp}(P,\mathit{post})+(1-p)\cdot \mathit{wp}(Q,\mathit{post})\\
\mathit{wp}(\while(G)\ \{P\}, \mathit{post})=\mu X.(\chi_G \cdot \mathit{wp}(P, X)+(1-\chi_G) \cdot \mathit{post})
\end{array}
\]
Here $\mathit{post}[x/E_{S}(E)]$ denotes the formula obtained by replacing free occurrences of $x$ in $\mathit{post}$ by the expectation of expression $E$ over the state space $S$. The least fixed point operator $\mu$ is defined over the domain of expectations~\cite{morgan1996probabilistic,McIver-book}, and it can be shown that $\langle \mathit{pre} \rangle\ P\ \langle \mathit{post} \rangle$ holds if and only if $\mathit{pre} \leq \mathit{wp}(P, \mathit{post})$. Thus, $\mathit{wp}(P, \mathit{post})$ is the greatest lower bound of precondition expectation of $P$ with respect to $\mathit{post}$, and we say $\mathit{wp}(P, \mathit{post})$ is the \textit{weakest pre-expectation of P w.\,r.\,t.\@ post}.

\subsection{Positivstellensatz}\label{subsec:pos}
Hilbert's Nullstellensatz is very important in algebra,
and its real version, known as Positivstellensatz, is crucial to our method.
First, some concepts are needed to introduce the theorem.

\begin{itemize}
\item	The set $P \subseteq \Real[\Xs]$ is a \textit{positive cone} if it satisfies:
	(i) If $a\in \Real[\Xs]$, then $a^2 \in P$, and
	(ii) $P$ is closed under addition and multiplication.
\item	The set $M \subseteq \Real[\Xs]$ is a \emph{multiplicative monoid with 0} if it satisfies:
	(i) $0, 1 \in M$, and
	(ii) $M$ is closed under multiplication.
\item	The set $I \subseteq \Real[\Xs]$ is an \textit{ideal} if it satisfies:
	(i) $0\in I$,
	(ii) $I$ is closed under addition, and
	(iii) If $a\in I$ and $b \in \mathbb{R}[\Xs]$, then $a \cdot b\in I$.
\end{itemize}
We are interested in finitely generated positive cones, multiplicative monoids with 0, and ideals.
Let $\mathcal{F} = \{f_1, \ldots, f_s \}$ be a finite set of polynomials.
We recall that
\begin{itemize}
\item	Any element in the positive cone generated by $\mathcal{F}$
	(i.\,e.\@, the smallest positive cone containing $\mathcal{F}$)
	is of the form
	\[
	\sum_{\alpha\in\{0,1\}^s} k_\alpha\mathcal{F}^\alpha\quad \text{where } k_\alpha \text{ is a sum of squares for all } \alpha\in\{0,1\}^s
	\]
	In the sum, $\alpha$ denotes an $s$-length vector with each element $0$ or $1$.
\item	Any nonzero element in the multiplicative monoid with 0 generated by $\mathcal{F}$
	is of the form $\mathcal{F}^{\alpha}$, where $\alpha=(\alpha_1,\ldots,\alpha_s) \in \mathbb{N}^s$.
\item	Any element in the ideal generated by $\mathcal{F}$ is of the form
	$k_1f_1+k_2f_2+\cdots+k_sf_s$,
	where $k_1,\ldots ,k_s\in \Real[\Xs]$.
\end{itemize}

The Positivstellensatz due to Stengle states that for
a system of real polynomial equalities and inequalities, either there
exists a solution, or there exists a certain polynomial which
guarantees that no solution exists.
\begin{theorem}[Stengle's Positivstellensatz~\cite{stengle1974nullstellensatz}]\label{thm:pos}
  Let $(f_j)_{j=1}^s, (g_k)_{k=1}^t,(h_l)_{l=1}^w$ be finite families of polynomials in $\Real[\Xs]$.
Denote by $P$ the positive cone generated by $(f_j)_{j=1}^s$,
by $M$ the multiplicative monoid with 0 generated by $(g_k)_{k=1}^t$,
and by $I$ the ideal generated by $(h_l)_{l=1}^w$.
Then the following are equivalent:
\begin{enumerate}
\item The set
	\begin{equation}\label{equ:pos}
	\left\{
	\zs\in \Real^n\Bigg|
	\begin{array}{cc}
	f_j(\zs)\geq 0,&j=1,\ldots, s\\
	g_k(\zs)\neq 0,&k=1,\ldots, t\\
	h_l(\zs) =0, &l=1,\ldots, w
	\end{array}\right\}
	\end{equation}
	is empty.
\item	There exist $f\in P, g\in M, h\in I$ such that $f+g^2+h=0$.

\end{enumerate}
\end{theorem}

\section{Problem formulation}\label{subsec:prob}
The question that concerns us here is to verify whether the loop sentence
\[
\langle \mathit{preE}\rangle\ \while (G)\ \{\mathit{body}\}\ \langle \mathit{postE} \rangle
\]
holds,
when given the pre-expectation $\mathit{preE}$, post-expectation $\mathit{postE}$,
a Boolean expression $G$, and a loop-free probabilistic program $\mathit{body}$.
One way to solve this problem is
to calculate the weakest pre-expectation $\mathit{wp}(\while(G, \{\mathit{body}\})$, $\mathit{postE})$
and to check whether it is not smaller than $\mathit{preE}$.
However, the weakest pre-expectation of a $\while$ statement requires a fixed-point computation, which is not trivial.
To avoid the fixed point, the problem can be solved through a quantitative loop invariant.

\begin{theorem}[\cite{gretz2013prinsys6}]\label{thm:1}
	Let $\mathit{preE}$ be a pre-expectation, $\mathit{postE}$ a post-expectation, $G$ a Boolean expression, and body a loop-free probabilistic program. To show
	\[
	\langle \mathit{preE}\rangle\ \while (G)\ \{\mathit{body}\}\ \langle \mathit{postE} \rangle,
	\]
	it suffices to find a loop invariant $I$ which is an expectation such that\\
        \begin{enumerate}
        \item
          (boundary) $\mathit{preE}\leq I$ and $I\cdot (1-\chi_G)\leq \mathit{postE}$;
        \item 	(invariant) $I\cdot \chi_G\leq \mathit{wp}(\mathit{body}, I)$;
        \item 	(soundness) the loop terminates with probability~1
		from any state that satisfies $G$, and
          \begin{enumerate}
          \item the number of iterations is finite, or
          \item $I$ is bounded from above by some fixed constant, or
          \item the expected value of $I\cdot \chi_G$ tends to zero as the number of iterations tends to infinity.
          \end{enumerate}
        \end{enumerate}
\end{theorem}
Since soundness of a loop invariant is not related to pre- and postconditions
and can be verified from its type before any specific invariants are found,
we focus on the boundary and invariant conditions in Theorem~\ref{thm:1}.
The soundness property is left to be verified manually in case studies.

For pre-expectation $\mathit{preE}$ and post-expectation $\mathit{postE}$,
the boundary and invariant conditions in Theorem~\ref{thm:1} provide the following requirements for a loop invariant $I$:
\begin{equation}\label{cons:1}
\begin{aligned}
\mathit{preE}&\leq I\\
I\cdot (1-\chi_{G})&\leq \mathit{postE}\\
I\cdot \chi_{G}&\leq \mathit{wp}(\mathit{body},I).
\end{aligned}
\end{equation}

The inequalities induced by the \emph{boundary} and \emph{invariant} conditions contain indicator functions,
which we find difficult to analyze if they appear on both sides.
So first we rewrite the expectations to a standard form.
For a Boolean expression $F$, we use $[F]$ to represent its integer value,
i.\,e.\@ $[F]=1$ if $F$ is true, and $[F]=0$ otherwise.
An expectation is in \textit{disjoint normal form} (DNF) if it is of the form
$$
f=[F_1]\cdot f_1+\cdots + [F_k]\cdot f_k
$$
where the $F_i$ are disjoint expressions,
which means any two of the expressions cannot be true simultaneously,
and the $f_i$ are polynomials.
\begin{lemma}[\cite{katoen2010linear9}]\label{lem:dnf2}
	Suppose $f=[F_1]\cdot f_1+\cdots+[F_k]\cdot f_k$ and $g=[G_1]\cdot g_1+\cdots+[G_l]\cdot g_l$ are expectations over $\Xs$ in DNF.
	Then, $f\leq g$ if and only if (pointwise)
	\begin{multline}\label{equ:dnf3}
	\bigwedge_{i = 1}^{k}\bigwedge_{j = 1}^l \Bigg[ F_i\mathrel{\wedge} G_j \Rightarrow f_i\leq g_j \Bigg] \\
	\mbox{} \mathrel{\wedge} \bigwedge_{i = 1}^k \Bigg[ F_i\mathrel{\wedge} \bigg( \bigwedge_{j = 1}^l \neg G_j \bigg) \Rightarrow f_i\leq 0 \Bigg] \\
	\mbox{} \mathrel{\wedge} \bigwedge_{j = 1}^l \Bigg[ \bigg( \bigwedge _{i=1}^k\neg F_i \bigg) \mathrel{\wedge} G_j \Rightarrow 0\leq g_j \Bigg]
	.
	\end{multline}
\end{lemma}
\begin{example}\label{exa:rm}
	Consider the following loop sentence for our running example:
	\[
	\langle xy-x^2\rangle\ z:=0;\ \while(0<x<y)\{x:=x+1\ [0.5]\ x:=x-1;z:=z+1;\}\ \langle z\rangle
	\]
	For this case, the following must hold for any loop invariant $I$.
	\begin{align*}
	xy-x^2&\leq I\\
	I\cdot [x\leq 0\mathrel{\vee} y\leq x]&\leq z\\
	I\cdot [0<x<y]&\leq 0.5\cdot I(x+1, y, z+1)+0.5\cdot I(x-1, y, z+1)
	\end{align*}
	By Lemma~\ref{lem:dnf2}, these requirements can be written as
	\begin{align}
	\mbox{}
	& \mathrel{\phantom{\Rightarrow}} xy-x^2 \leq I \mathrel{\wedge} \mbox{} \label{equ:3}\\
	x\leq 0\mathrel{\vee} y\leq x &\Rightarrow I\leq z \mathrel{\wedge} \mbox{} \label{equ:4}\\
	0<x<y &\Rightarrow 0\leq z \mathrel{\wedge} \mbox{} \label{equ:5}\\
	0<x<y &\Rightarrow I\leq 0.5\cdot I(x+1, y, z+1)+0.5\cdot I(x-1, y, z+1) \mathrel{\wedge} \mbox{} \label{equ:6}\\
	x\leq 0\mathrel{\vee} y\leq x &\Rightarrow 0\leq 0.5\cdot I(x+1, y, z+1)+0.5\cdot I(x-1, y, z+1) \label{equ:7}
	\end{align}
\end{example}
The program in this example originally served as a running example in~\cite{chen2015counterexample}.
There, after transforming the constraints into the form above,
Lagrange interpolation is applied to synthesize the coefficients in the template.
In our approach, we check the correctness of each conjunct in (\ref{equ:3}--\ref{equ:7})
by checking the nonnegativity of the polynomial on the right side over a semialgebraic set related to polynomials on the left side.
In this way, we can use the Positivstellensatz to synthesize the coefficients.

\section{Constraint Solving by Semidefinite Programming}\label{sec:sdp}
Our aim is to synthesize coefficients for the fixed invariant template for simple
(Subsection~\ref{subsec:sm1}) and nested 
(Subsection~\ref{subsec:sm2}) programs. Checking the validity of constraints can be transformed into checking the emptiness of
a semialgebraic set. Then, we show that the
emptiness problem can be turned into sum-of-squares constraints using Stengle's Positivstellensatz.

\paragraph{Our Approach in a Nutshell.}
For a given polynomial template as a candidate quantitative loop invariant,
it needs to satisfy \emph{boundary} and \emph{invariant} conditions.
Our goal is to synthesize the coefficients in the template.
These conditions describe a semialgebraic set,
and the satisfiability of the constraints is equivalent
to the non-emptiness of the corresponding semialgebraic set.
Applying the Po\-si\-tiv\-stel\-len\-satz (see Section~\ref{subsec:pos}),
we will transform the problem to an equivalent semidefinite programming problem
using Lemma~\ref{lem:equ}.
Existing efficient solvers can be used to solve the problem.
A more efficient yet sufficient way is to transform the problem into a \emph{sum-of-squares problem} (see Appendix~\ref{subsec:sos})
using Lemma~\ref{lem:alg} and then to solve it by semidefinite programming.
After having synthesized the coefficients of the template,
we verify whether they are valid.
In case of a negative answer, which may happen due to floating-point errors,
some refinements can be made by adding further constraints, which is described in Section~\ref{sec:cons}.
If the problem is still unsolved,
we try raising the maximum degree of the template and reiterate the procedure.

\subsection{Synthesis Algorithm for Simple Loop Programs}\label{subsec:sm1}

Now we are ready for the transformation method.
Each conjunct obtained in Lemma~\ref{lem:dnf2} is of the form $ F \Rightarrow G$,
where $F$ is a quantifier-free formula constructed from (in)equalities between polynomials in $\Real^{\leq d}[\Xs]$,
and $G$ is of the form $f\leq g$, $f\leq 0$ or $0\leq g$,
with $f,g \in \Real^{\leq d}[\Xs]$.
If $F$ contains negations, we use De Morgan's laws to eliminate them.
If there is a disjunction in $F$,
we split the constraints into sub-constraints
as $\varphi \mathrel{\vee} \psi\Rightarrow \chi$ is equivalent to $(\varphi \Rightarrow \chi) \mathrel{\wedge} (\psi \Rightarrow \chi)$.
After these simplifications,
$F \Rightarrow G$ can be written in the form
$
\bigwedge_i( f_i \unrhd_i 0) \Rightarrow g \geq 0
$
where $\mathord{\unrhd_i} \in\{\mathord{\geq}, \mathord{=}\}$.
Observe that a constraint $\bigwedge_i( f_i \unrhd_i 0) \Rightarrow g \geq 0$ is satisfied
if and only if the set  $\{x|f_i(x)\unrhd_i 0 \text{ for all } i \text{; } {-g(x)}\geq 0 \text{; and } g(x)\neq 0\}$ is empty.
In this way, we transform our constraint into the form required by Theorem~\ref{thm:pos}.

Summarizing, Constraint~\eqref{cons:1} (the boundary and invariant conditions of Theorem~\ref{thm:1}) is satisfied
if and only if all semialgebraic sets created using the procedure above are empty.
Now we are ready to transform this constraint to an SDP problem. 

\begin{lemma}[\cite{parrilo2003semidefinite,dai2013generating}]\label{lem:equ}
	The emptiness of \eqref{equ:pos} is equivalent to the feasibility of an SDP problem.
\end{lemma}	

See Appendix~\ref{subsec:prfequ} for a constructive proof.
Although the transformation in Lemma~\ref{lem:equ} is effective,
it is complicated in practice.
In the following lemma we present a simpler yet sufficient procedure.  	

\begin{lemma} \label{lem:alg}
	The following statements hold
	(with $\mathord{\unrhd_i} \in\{\mathord{\geq}, \mathord{=}\}$):
	\begin{enumerate}
	\item	$f(\Xs)\geq 0 \Rightarrow g(\Xs) \geq 0$ holds
		if $g(\Xs)-u\cdot f(\Xs)$ is a sum of squares
		for some $u \in \Real_{\geq 0}$.
	\item	$f(\Xs)= 0 \Rightarrow g(\Xs) \geq 0$ holds
		if $g(\Xs)-v\cdot f(\Xs)$ is a sum of squares
		for some $v \in \Real$.
	\item	$f_1(\Xs) \unrhd_1 0 \mathrel{\wedge} f_2(\Xs) \unrhd_2 0 \Rightarrow g(\Xs) \geq 0$ holds
		if $g(\Xs)-r_1\cdot f_1(\Xs) - r_2\cdot f_2(\Xs)$ is a sum of squares
		for some $r_1, r_2 \in \Real$;
		if $\unrhd_i$ is $\geq$, it is additionally required that $r_i \geq 0$.
		\label{item:alg:conjunction}
	\end{enumerate}
\end{lemma}
The proof is in Appendix~\ref{subsec:prfalg}.
Note that Item~(\ref{item:alg:conjunction}) is one of the possible sufficient relaxations;
more general relaxations can be obtained by adding a cross product $r_{12}f_1(\Xs)f_2(\Xs)$ and squares of the $f_i(\Xs)$.

\begin{example}\label{exa:sos}
	Applying the above procedure,
	Constraint~\eqref{equ:4} in Example~\ref{exa:rm} is split into
	$(x\leq 0 \Rightarrow I\leq z) \mathrel{\wedge} (y\leq x \Rightarrow I \leq z)$
	and then normalized to
	$(-x\geq 0 \Rightarrow z-I\geq 0) \mathrel{\wedge} (x-y\geq 0 \Rightarrow z-I \geq 0)$.
	This holds if $z-I+u_1 x$ is a SOS for some $u_1 \in \Real_{\geq 0}$
	and $z-I+ u_2 (y-x)$ is a SOS for some $u_2 \in \Real_{\geq 0}$.
	The other constraints can be handled in a similar way.
\end{example}

After applying the Positivstellensatz and Lemma~\ref{lem:equ},
template coefficients for the loop invariant can be synthesized efficiently by semidefinite programming.
The corresponding technique is introduced in Appendix~\ref{app:sdp}.

\begin{algorithm}[thb]
	\caption{Loop Invariant Generation with Refinement\label{arg1}}
	\textbf{Input:} $\mathit{sentence} := \langle \mathit{preE}\rangle\ \while (G) \{\mathit{body}\}\ \langle \mathit{postE}\rangle$
			with program variables $\Xs$ \\
	\textbf{Output:} a loop invariant satisfying the boundary and invariant conditions \\[-\baselineskip]
	\begin{algorithmic}[1]
		\LOOP
		\STATE $d := 2$
		\STATE Choose a template for $I \in \Real^{\leq d}[\Xs]$
		\STATE Let $f$ be Constraint~\eqref{cons:1}, i.\,e.\@ the boundary and invariant conditions from Theorem~\ref{thm:1}, for $\mathit{sentence}$
		\STATE Let $\mathit{constraints}$ be the SDP problem equivalent to $f$ according to Lemma~\ref{lem:equ}
			\label{alg:find_SDP_problem}
		\WHILE{$\mathit{constraints}$ is feasible}\label{alg:feasible}
		\STATE Set the coefficients in the template for $I$
		\STATE Round the coefficients of $I$ into rational numbers 
		\IF{$I$ satisfies the boundary and invariant conditions}
		\STATE Output $I$ and terminate
		\ENDIF
		\STATE Refine $\mathit{constraints}$
		\ENDWHILE
		\STATE $d := d+2$
		\ENDLOOP
	\end{algorithmic}
\end{algorithm}
We summarize our approach in Algorithm~\ref{arg1}.
The aim is to synthesize the coefficients of template $I$.
The terms in $I$ are all terms with degree $\leq d$ in the multiplicative monoid generated by $\Xs$.
%
Algorithm~\ref{arg1} is semi-complete in the sense that
it will generate an invariant if there exists one.
Its termination is guaranteed in principle by Theorem~\ref{thm:pos} and the equivalence between SOS and SDP in lemma~\ref{lem:equ},
though due to numerical errors, the algorithm may fail to find $I$ in practice.

In practice, Lemma~\ref{lem:alg} is often used instead of Lemma~\ref{lem:equ} for efficiency.
Step~\ref{alg:find_SDP_problem} in Algorithm~\ref{arg1} is replaced by:
``Let $\mathit{constraints}$ be the relaxation of $f$ to an SOS problem according to Lemma~\ref{lem:alg}'';
this can be translated to an equivalent SDP problem,
which is simpler than the direct translation of Lemma~\ref{lem:equ},
using the technique of Appendix~\ref{subsec:sos}.

\begin{example}\label{exa:ful}
We extend Example~\ref{exa:rm} using Lemma~\ref{lem:alg}.
To illustrate our solution method,
we choose Constraints~\eqref{equ:3}, \eqref{equ:4}, and \eqref{equ:6}.
The initial condition $z=0$ is not included in these constraints,
so \eqref{equ:3} needs to be refined to $z=0\Rightarrow xy-x^2\leq I$.
	
	First, we set a template for $I$. Assume $I$ as a quadratic polynomial with three variables $x, y, z$:
	$$
	I=c_0+c_1x+c_2y+c_3z+c_{11}x^2+c_{12}xy+c_{13}xz+c_{22} y^2+c_{23}yz+c_{33}z^2
	$$
	where $c_0, \ldots, c_{33} \in \Real$ are coefficients that remains to be determined.
	
	For Constraint~\eqref{equ:3} with initial constraint $z=0$, we get the following corresponding constraint:
	$$
	I-(xy-x^2) - v\cdot z\geq 0 \eqno{(\ref{equ:3}')}
	$$
	
	For \eqref{equ:4}, the antecedens is a conjunction of two constraints.
	As in Example~\ref{exa:sos}, \eqref{equ:4} is split into two constraints and transformed into
	\begin{align*}
	z-I + u_1\cdot x & \geq 0\quad \textrm{and} \\
	z-I - u_2\cdot (x-y) & \geq 0   \tag{\ref{equ:4}$'$}
	\end{align*}
	
	For \eqref{equ:6}, the constraint $0<x<y$ needs to be split into two inequalities $x > 0 \mathrel{\wedge} y-x > 0$.
	Similarly to \eqref{equ:4},
	we transform \eqref{equ:6} to
	\[
	0.5\cdot I(x+1, y, z+1)+0.5\cdot I(x-1, y, z+1)-I
	-u_3\cdot x-u_4\cdot(y-x)\geq 0 \eqno{(\ref{equ:6}')}
	\]
	In this way the example can be transformed into an SDP problem with constraints~(\ref{equ:3}$'$), (\ref{equ:4}$'$), and (\ref{equ:6}$'$),
	and positivity constraints on the multipliers $u_1\geq 0, \ldots, u_4\geq 0$.
	(For $v$, an arbitrary real value is allowed.)
	Then the resulting SDP problem can be submitted to any SOS solver.
	
	The result using solver SeDuMi~\cite{SeduMi} is shown below.
	\begin{multline*}
	I = -7.1097 \cdot 10^{-10} - 3.8818 \cdot 10^{-10} x - 0.4939 \cdot 10^{-10} y + z - x^2 + xy + \\
	2.7965 \cdot 10^{-10} xz + 0.97208 \cdot 10^{-10} y^2 + 4.4656 \cdot 10^{-10} yz - 0.28694 \cdot 10^{-10} z^2
	\end{multline*}
	If we ignore the amounts smaller than the order of magnitude of $10^{-6}$,
	we get $I = z - x^2 + xy$.
	This $I$ satisfies all constraints including \eqref{equ:5} and \eqref{equ:7}, so it is correct.
\end{example}

\subsection{Synthesis Algorithm for Nested Loop Programs}\label{subsec:sm2}
We are now turning to programs containing nested loops.
To simplify our discussion, we assume the program only contains a single, terminating inner loop,
i.\,e.\@ it can be written as
\begin{align*}
P & =\while(G)\{\mathit{body}\} \\
& =\while(G)\{\mathit{body1};\ \while(G_\mathrm{inn}) \{\mathit{body}_\mathrm{inn}\};\ \mathit{body2} \}
\end{align*}
where $\mathit{body1}$, $\mathit{body}_\mathrm{inn}$, and $\mathit{body2}$ are loop-free program fragments.
(If the inner loop is placed within an $\progif$ statement, one can transform it to the above form by strengthening $G$.)
For a given $\mathit{preE}$ and $\mathit{postE}$, we need to verify
whether there exists an invariant $I$ that satisfies Constraint~\eqref{cons:1}
(the boundary and invariant conditions of Theorem~\ref{thm:1}).
We denote the inner loop by $P_\mathrm{inn}=\while(G_\mathrm{inn}) \{\mathit{body_{inn}}\}$.

For such a program, the main difficulty is how to deal with $\mathit{wp}(\mathit{body},I)$ in Constraint~\eqref{cons:1}.
We propose a method here that takes the inner and outer iteration into consideration together
and uses the verified pre-expectation of the inner loop to relax the constraint.

Fix templates for the polynomial invariants:
$I$ for the outer loop and $I_\mathrm{inn}$ for the inner loop $P_\mathrm{inn}$,
both with degree $d$.
Since $\mathit{body2}$ is loop-free, it is easy to obtain $\tilde{I} := \mathit{wp}(\mathit{body2}, I)$.
We use $\tilde{I}$ as post-expectation $\mathit{postE}_\mathrm{inn}$ for the inner loop.
Note that \eqref{cons:1} for the inner loop requires $\mathit{preE}_\mathrm{inn} \leq I_\mathrm{inn}$,
so we can use the template $I_\mathrm{inn}$ also as template for $\mathit{preE}_\mathrm{inn}$.
Then the constraints for loop invariant $I$ are
\begin{equation}\label{cons:2}
\begin{aligned}
\mathit{preE}&\leq I\\
I\cdot [1-\chi_{G}]&\leq \mathit{postE}\\
I\cdot \chi_{G}&\leq \widetilde{\mathit{wp}}(\mathit{body},I)=
\mathit{wp}(\mathit{body1}, \mathit{preE_\mathrm{inn}})\\
\mathit{preE}_\mathrm{inn}&= I_\mathrm{inn}\\
I_\mathrm{inn}\cdot [1-\chi_{G_\mathrm{inn}}] &\leq \mathit{postE}_\mathrm{inn}=\mathit{wp}(\mathit{body2}, I)\\
I_\mathrm{inn}\cdot \chi_{G_\mathrm{inn}}&\leq \mathit{wp}(\mathit{body}_\mathrm{inn},I_\mathrm{inn})
\end{aligned}
\end{equation}
The first three equations are almost Constraint~\eqref{cons:1} for the outer loop,
except that $\widetilde{\mathit{wp}}$ is the strengthening of the weakest pre-expectation
using $\mathit{preE}_\mathrm{inn} = I_\mathrm{inn}$ in the $\mathit{wp}$-calculation
instead of $\mathit{wp}(P_\mathrm{inn},\tilde{I})$.
The last three equations are Constraint~\eqref{cons:1} for the inner loop,
except that we require equality in $\mathit{preE}_\mathrm{inn} \leq I_\mathrm{inn}$.

Then we have the following lemma.
\begin{lemma}\label{lem:nest}
	An invariant $I$ that satisfies Constraint~\eqref{cons:2} also satisfies \eqref{cons:1},
	therefore it is a loop invariant for program $P$.
\end{lemma}
See Appendix~\ref{subsec:prfnest} for the proof.

\subsection{Handling Numerical Error}\label{sec:cons}
In practice, it sometimes happens
that numerical errors lead to wrong or trivial coefficients in the templates.
We suggest several methods to refine the constraints and avoid these errors.

Due to the inaccuracy of floating-point calculations,
it is hard for a software to check equations and inequalities like $x = 0$ or $x\neq 0$.
A common trick to avoid this problem is to turn the equality constraint into $x \geq 0 \mathrel{\wedge} x \leq 0$.
As for inequalities, taking $x\neq 0$ as an example,
a way to solve the problem is adding a new variable $y$ to transform the constraint into $xy \geq 1$,
since $xy \geq 1$ implies $x \neq 0$ for any value of $y$.
The new constraints are in the form required by Theorem~\ref{thm:pos}.


Numerical errors may also lead to an unsound invariant:
we may get some coefficients with a small magnitude,
which often result from floating-point inaccuracies.
A common solution for this problem is to ignore those small numbers, usually smaller than $10^{-6}$ in practice.
In Example~\ref{exa:ful}, eliminating the terms with a small order of magnitude was successful,
but we cannot be sure whether the resulting invariant is correct
if the remaining coefficients are approximate.
We propose to check the soundness of such solutions symbolically as follows.
Checking whether the generated invariant satisfies Constraint~\eqref{cons:1}
is a special case of quantifier elimination $\forall \xs\in\Real^n,\ f(\xs)\geq 0$.
Such problem can be solved efficiently using an improved Cylindrical Algebraic Decomposition (CAD) algorithm implemented in~\cite{han2016proving}.
In our experiments in Section~\ref{sec:exp}, the found solutions are obtained by ignoring small numbers,
and we verified they are correct by running CAD in a separate tool.

If the invariant still violates some of the constraints,
we can try to strengthen the constraint (e.\,g., change $x \geq 0$ to $x \geq 0.1$)
and repeat our method.

\section{Experimental Results}\label{sec:exp}

We have implemented a prototype in Python to test our technique.
We call the MATLAB toolbox YALMIP~\cite{lofberg2004yalmip} with the SeDuMi solver~\cite{SeduMi}
to solve the SDP feasibility problem.
We use the math software Maple to verify the correctness of the constraints through CAD.
The experiments were done on a computer with Intel(R) Core(TM) i7-4710HQ CPU and 16\,GiB of RAM.
The operating system is Window~7 (32bit).
Constraint refinement cannot be handled automatically in the current version,
but we plan to add it together with projection for rounding solutions in a future version.

Our prototype and the detailed experimental results can be found at \url{http://iscasmc.ios.ac.cn/ppsdp}.
For each probabilistic program, we give the description of the while loop with pre- and post-expectations in Table~\ref{tab:mainresult} and Appendix~\ref{subsec:exm}.
The annotated pre-expectation serves as an exact estimate of the annotated post-expectation at the
entrance of the loop. We apply the method to several different types of examples.
A summary of the results is shown in Table~\ref{result}.
The first eleven probabilistic programs are benchmarks taken from paper~\cite{chen2015counterexample},
thus we skip the detailed descriptions of them.
We have further constructed three case studies to illustrate continuous distributions,
polynomial probabilistic programs and nested loop programs.
The details of these examples are included in Appendix~\ref{subsec:exm}.
We ran CAD in Maple manually to verify the feasability of the generated invariants.

As we can see from Table~\ref{tab:mainresult}, the
running time of our method is within one second. There are some notes when calculating the examples. We relax the loop condition $z\neq 0$ in example geo2 into $z\geq 0.5$. Also in the fair coin example, we relax the loop condition $x\neq y$ into $x-y\geq 0.5\vee y-x\geq 0.5$. Since variables in those two examples are integers, the relaxation is sound.

\begin{table}[htb]\label{tab:mainresult}
	\caption{%
The column ``Name'' shows the name of each experiment.
The annotated pre- and post-expectations are shown in the columns ``$\mathit{preE}$'' and ``$\mathit{postE}$''.
The inferred quantitative loop invariant for each example is given in the column ``Invariant''.
The column ``Time'' gives the running time needed of our tool:
the first one is the total running time, and the second one is the time used in the SeDuMi solver.}
	\label{result}
	\centering
	\begin{tabular}{|m{45pt}|m{55pt}|c|m{150pt}|c|}
		\hline
		Name&$\mathit{preE}$&$\mathit{postE}$&\multicolumn{1}{c|}{Invariant}&Time (s)\\\hline
		ruin&$xy-x^2$&$z$&$z+xy-x^2$&$0.4/0.3$\\\hline
		bin1&$x+\frac{1}{4}ny$&$x$&$x+\frac{1}{4}ny$&$0.4/0.2$\\\hline
		bin2&$\frac{1}{8}n^2-\frac{1}{8}n+\frac{3}{4}ny$&$x$&$x+\frac{1}{8}n^2-\frac{1}{8}n+\frac{3}{4}ny$&$0.7/0.3$\\\hline
		bin3&$\frac{1}{8}n^2-\frac{1}{8}n+\frac{3}{4}ny^2$&$x$&$x-0.0057n-0.0014x^2+0.1763xn+712.909n^2+0.0014x^2n+0.4114xn^2+0.4188ny^2-0.0178n^3$&$0.7/0.3$\\\hline
		geo&$x+3zy$&$x$&$x+3zy$&$0.2/0.2$\\\hline
		geo2&$x+\frac{15}{2}$&$x$&$x+30.2312y+3.4699z-12.6648y^2-44.6591yz-35.5112z^2-22.8807$&$0.2/0.1$\\\hline
		sum&$\frac{1}{4}n^2+\frac{1}{4}n$&$x$&$x+\frac{1}{4}n^2+\frac{1}{4}n$&$0.3/0.1$\\\hline
		prod&$\frac{1}{4}n^2-\frac{1}{4}n$&$xy$&$-\frac{1}{4}n+xy+\frac{1}{2}xn+\frac{1}{2}yn+\frac{1}{4}n^2$&$0.3/0.1$\\\hline
		fair coin1&$\frac{1}{2}-\frac{1}{2}x$&$1-x+xy$&$0.7130-0.5622x+0.3364y+0.8564n-1.2740x^2+07610xy-1.4572xn-1.2208y^2+1.4572yn-0.1366n^2$&$0.2/0.1$\\\hline
		fair coin2&$\frac{1}{2}-\frac{1}{2}y$&$x+xy$&$1.1941+1.6157x+0.6387y+7.9774n-14.6705x^2+9.7904xy-14.9948xn-14.6457y^2+14.9948yn-1.4058n^2$&$0.3/0.1$\\\hline
		fair coin3&$\frac{8}{3}-\frac{8}{3}x-\frac{8}{3}y+\frac{1}{3}n$&$n$&$6.0556+2.5964x+3.2468y+39.2052n-69.9038x^2+44.0224xy-72.1408xn-69.8067y^2+72.1408yn-6.7632n^2$&$0.2/0.1$\\\hline
		simple &&&&\\ perceptron&$-2b$&$n$&$n-2b$&0.3/0.1\\\hline
		airplane &&&&\\ delay&$106.548x$&$h$&$106.548x-106.548n+h$&$0.4/0.2$\\\hline
		airplane &&&&\\ delay2&$282.507x$&$h$&$282.507(x-n)+h$&$0.5/0.2$\\\hline
		nested loop&$20(m-x)$&$k$&$k+20(m-x)$&$1.6/1.1$\\\hline
	\end{tabular}
\end{table}

\subsection{Evaluation}\label{subsec:eval}
Other approaches to compute loop invariants in probabilistic programs are
the Lagrange Interpolation Prototype (LIP) in~\cite{chen2015counterexample},
the tool for martingales (TM) in~\cite{chakarov2013probabilistic}
and \textsc{Prinsys} in~\cite{gretz2013prinsys6}.
The tools are executed on the same computer, LIP and TM under Linux and the other two under Windows.
In Table~\ref{tab:comparison},
we compare the features supported by the four tools.

\begin{table}[tbh]
	\centering
	\caption{\label{tab:comparison} Comparison of the features supported by 4 tools}
	\begin{tabular}{|m{90pt}|c|c|c|c|}
		\hline
		&\textsc{Prinsys}&LIP&TM&Our tool\\\hline
		Type of Program&Linear&Cubic&Linear&Polynomial\\\hline
		Type of Invariant&Linear&Polynomial&Linear&Polynomial\\\hline
		Computation Method&Symbolic&Symbolic&Numerical&Numerical\\\hline
		Distribution         of Variable&Discrete&Discrete&Continuous&Continuous\\\hline
	\end{tabular}
\end{table}

We have tested the examples in Table~\ref{result} on these four tools.
\textsc{Prinsys} takes the longest time and fails to verify any of non-linear examples presented.
LIP fails to verify any examples that include a continuous variable or have a degree larger than 3;
additionally it is always about 10~times slower than our tool.
TM fails to verify examples ruin, bin3 and geo directly.
We observe that it cannot treats constraints of the form $x=y$ or $x\neq y$
(where $x$ and $y$ might be variables or constants).
However, by transforming $x=y$ into $x\geq y \wedge y\geq x$,
TM can synthesize a supermartingale for the program.
Also, it cannot verify the simple perceptron, as it is a non-linear program.
Furthermore, TM cannot deal with nested loop programs.

We now consider the parametric linear program in Section~\ref{sec:pe}.
Table~\ref{tab:compare} gives a comparison of time consumption of the main technical step in our prototype.
The number of constraints grows with the number of variables in our approach, similarly with the running time. Some more experiments on the number of variables and maximum degree of polynomials are included in Appendix~\ref{sec:pe}.

\begin{table}[hbt]
	\centering
	\caption{\label{tab:compare}Comparison of running time (in seconds) of the parameterized linear example}
	\begin{tabular}{|l|c|c|c|c|c|c|}
		\hline
		Number of variables&$\ n=15\ $ &$\ n=20\ $ &$\ n=25\ $ & $\ n=30\ $ & $\ n=35\ $ & $\ n=40\ $ \\\hline
		Solver time of our tool&$0.41$&$1.30$&$2.44$&$8.30$&20.56&46.62\\\hline
	\end{tabular}
\end{table}

\section{Conclusion}\label{sec:conclusion}
In this paper, we propose a method
to synthesize polynomial quantitative invariants for recursive probabilistic programs
by semialgebraic programming via a Positivstellensatz.
First, a polynomial template is fixed whose coefficients remain to be determined.
The loop and its pre- and post-expectation can be transformed into a semialgebraic set,
of which the emptiness can be decided by finding a counterexample satisfying the condition of the Positivstellensatz.
Semidefinite programming provides an efficient way to synthesize such a counterexample.
The method can be applied to polynomial programs containing continuous or discrete variables,
including those with nested loops.
When numerical errors prevent finding a loop invariant polynomial right away,
we currently can correct them \emph{ad hoc}
(by deleting terms with very small coefficients and sometimes strengthening the constraints),
but we would like to develop a more systematic treatment.

As future improvements, we are considering improvements in numerical error handling.
A better approximation can be found by projecting $\tilde{I}(x)$ onto a rational subspace defined by SDP constraints~
\cite{peyrl2008computing,kaltofen2012exact}.
There are also acceleration methods for different types of probabilistic programs:
For \emph{linear} programs, Handelman's Positivstellensatz describes a faster way to synthesize SOS constraints,
and for \emph{Archimedean} programs, \cite{dai2013generating} describes a faster way to apply Stengle's Positivstellensatz.

\bibliographystyle{splncs03}
\bibliography{myref}

\clearpage

\noindent
The following appendices are added for the convenience of the reviewers.
They will become part of a technical report accompanying our publication, once accepted.
We appreciate the reviewers' consideration.---The authors.

\appendix

\section{Sum-of-Squares Problems}\label{subsec:sos}

The set of sum-of-squares polynomials is a proper subset of the nonnegative polynomials with good algebraic properties,
allowing efficient calculations.
The ``Gram matrix'' method~\cite{choi1995sums} is a way to decompose a polynomial into a sum of squares using semidefinite programming.

Consider a polynomial $f = \sum_{|\alpha|\leq d}c_{\alpha}\Xs^{\alpha} \in \Real^{\leq d}[\Xs]$ with degree at most $d$.
$f$ is a sum of squares
if and only if it has a representation $f=\Xs A \Xs^T = \sum_{\alpha, \beta\in\Delta}a_{\alpha\beta}\Xs^{\alpha}\Xs^{\beta}$,
where the $|\Delta|\times|\Delta|$ matrix $A=(a_{\alpha\beta})_{\alpha, \beta\in\Delta}$ is positive semidefinite.
So checking whether $f$ is a sum of squares is equivalent to solving the following constraint problem:
\begin{equation}
\left\{
\begin{array}{ll}\label{eq1}
&c_0=a_{00},\\
&c_{\gamma}=\sum_{\alpha+\beta=\gamma}a_{\alpha\beta}\textrm{ for }\gamma\neq 0\\
&\textrm{and }A=(a_{\alpha\beta})_{\alpha, \beta\in\Delta}\textrm{ is positive semidefinite}
\end{array}\right.
\end{equation}
The above problem can be solved using a semidefinite program;
we refer to Appendix~\ref{app:sdp} for details.
Semidefinite programs can be efficiently solved both in theory and in practice
and have seen active research in recent years.
Some of the tools being used are SeDuMi~\cite{SeduMi} and SDPT3~\cite{sdpt3}.

\section{Semidefinite Programming}\label{app:sdp}
A semidefinite program can be seen as a generalization of a linear program
where the constraints are described by a cone of positive semidefinite matrices.

We use $\mathcal{S}^n$ to denote all real $n\times n$ symmetric
matrices. Then for a matrix $A\in \mathcal{S}^n$, $A$ is
\textit{positive semidefinite} if all eigenvalues of $A$ are $\geq 0$
(one can find more about positive semidefinite matrix in any book
about linear algebra such as~\cite{meyer2000matrix}). For matrices $A$
and $B$ in $\mathcal{S}^n$,
we write $A\succeq B$ if and only if $A-B$ is positive semidefinite.

We use $\langle \mathord{\,\cdot\,} , \mathord{\,\cdot\,} \rangle$ to denote the scalar product of two matrices or vectors,
i.\,e.\@ for $A=(a_{ij}), B=(b_{ij}) \in \mathcal{S}^n$ and $x, y\in\Real^n$
$$
\langle A,B\rangle:=Tr(A^TB)=\sum_{i,j=1}^n a_{ij}b_{ij}
$$
$$
\langle x,y\rangle:=x^Ty
$$
A semidefinite program is defined as follows:
\begin{equation}\label{eq:sdp}
\begin{array}{ll}
\text{minimize}  &\langle C, X \rangle\\
\text{subject to}&\langle A_i, X\rangle = b_i \text{ for } i=1,\ldots,k\\
                 &X\succeq 0
\end{array}
\end{equation}
where $X \in \mathcal{S}^n$ is the decision variable, $b_i \in \Real$ and $C$, $A_i \in \mathcal{S}^n$ are given symmetric matrices.

Sum-of-squares problem~\eqref{eq1} can be transformed into an SDP problem as follows.
Let $X=(a_{\alpha\beta})_{\alpha,\beta\in\Delta}$, $A_i$ (for $i \in \Delta$) be the symmetric matrix
whose $(\alpha,\beta)$ entry is 1 if $\alpha+\beta=i$
and 0 otherwise,
and let $b_i=c_i$.
In this way, \eqref{eq1} is transformed into the form \eqref{eq:sdp} without objective.

\section{Proofs of Lemmas}
\subsection{Proof of Lemma~\ref{lem:equ}}\label{subsec:prfequ}
\paragraph{\bfseries Lemma~\ref{lem:equ} (\cite{parrilo2003semidefinite,dai2013generating}).}
{\itshape
	The emptiness of \eqref{equ:pos} is equivalent to the feasibility of an SDP problem.
}

\begin{proof}
	We use the notations from Theorem~\ref{thm:pos} in this proof.
	The emptiness of \eqref{equ:pos} is equivalent to the existence of a solution for equation $f+g^2+h=0$ with $f\in P$, $g\in M$, $h\in I$ due to Theorem~\ref{thm:pos}.
	
	Note that $f$, $g$ and $h$ can be represented as
	\begin{itemize}
		\item $f = \sum_{\alpha\in\{0,1\}^s} u_\alpha\mathcal{F}^\alpha,$
		where $u_\alpha$ is a sum of squares for all $\alpha\in\{0,1\}^s$ and $\mathcal{F}=\{f_1,\ldots,f_s\}$, and
		\item	$g = \mathcal{G}^{\alpha}$ with $\mathcal{G}=\{g_1,\ldots,g_t\}$, and
		\item	$h=v_1h_1+v_2h_2+\cdots+v_wh_w$,
		where $v_1,\ldots ,v_w\in \Real[\Xs]$.
	\end{itemize}
	Fix a maximal degree $d$.
	Then we only consider $\mathcal{F}^\alpha$ and $\mathcal{G}^{\alpha}$ with $|\alpha|\leq d$.
	We set templates with degree $d$ for $u_\alpha$ and $v_0, \ldots, v_w$ and treat their coefficients
	as parameters.
	Every polynomial $v_i$ can be presented as the difference of two SOS polynomials:
	$v_i = \frac{(v_i+1)^2-(v_i-1)^2}{4}$.
	Then $f+g^2+h = \sum_{i=1}^{l}\delta_i p_i$,
	where $l$ is some integer,
	$p_i$ is one of $f_i$, $g_i$, $h_i$, or $-h_i$, and $\delta_i$ is a SOS.
	Then the equation $f+g^2+h=0$ can be transformed into a set of equations by merging coefficients of each $\Xs^\alpha$ with maximal degree $2d$.
	One can formulate additional constraints that $\delta_\alpha$ needs to be a SOS
	and the coefficient matrix $A_\alpha$ with $\delta_\alpha = \Xs A_\alpha \Xs^T$ to be positive semidefinite.
	The equation set with constraints can be transformed into an SDP problem of the form in Appendix~\ref{app:sdp}.
\end{proof}
\subsection{Proof of Lemma~\ref{lem:alg}}\label{subsec:prfalg}
\begin{proof}
	We only prove (1) and (2) here, (3) is a straightforward extension of them
	by analogy.
	
	For (1), assume $h(\Xs)=g(\Xs)- u\cdot f(\Xs)$ is an SOS for some $u\in \Real_{\geq 0}$ and $f(\Xs)\geq 0$. Since $h(\Xs)\geq 0$, $g(\Xs)=h(\Xs)+u\cdot f(\Xs)\geq 0$.
	
	For (2), assume $h(\Xs)=g(\Xs)- v\cdot f(\Xs)$ is an SOS for arbitrary $v\in\Real$ and $f(\Xs)= 0$. Then $g(\Xs)=h(\Xs)+v\cdot f(\Xs)=h(\Xs)\geq 0$. 
%
\end{proof}
By the method indicated above Lemma~\ref{lem:alg}, one can easily find that a constraint of the form $f_1(\Xs) \geq 0 \vee f_2(\Xs) \geq 0 \Rightarrow g(\Xs) \geq 0$ can be translated to two SOS problems.
\subsection{Proof of Lemma~\ref{lem:nest}}\label{subsec:prfnest}
\paragraph{\bfseries Lemma~\ref{lem:nest}.}
{\itshape
	An invariant $I$ that satisfies Constraint~\eqref{cons:2} also satisfies \eqref{cons:1},
	therefore it is a loop invariant for program $P$.
}
\begin{proof}
	The first two inequalities of \eqref{cons:1} are literally the same as the first two of \eqref{cons:2}.
	To prove the remaining inequality of \eqref{cons:1},
	we assume that the inner loop terminates.
	(The verification of soundness is not a part of our algorithm.)
	From Theorem~\ref{thm:1} applied to the last three (in)equalities in \eqref{cons:2},
	we immediately get that $\mathit{preE}_\mathrm{inn} \leq \mathit{wp}( P_\mathrm{inn}, \mathit{postE}_\mathrm{inn})$.
	From this we have:
	$$
	\begin{array}{rll}
	I \cdot \chi_G \leq \widetilde{\mathit{wp}}(\mathit{body}, I)
	& = \mathit{wp}(\mathit{body1}, \mathit{preE}_\mathrm{inn}) \\
	& \leq \mathit{wp}(\mathit{body1}, \mathit{wp}( P_\mathrm{inn}, \mathit{postE}_\mathrm{inn})) \\
	& \leq \mathit{wp}(\mathit{body1}, \mathit{wp}( P_\mathrm{inn}, \mathit{wp}(\mathit{body2}, I))) \\
	& = \mathit{wp} (\mathit{body1} ; P_\mathrm{inn} ; \mathit{body2}, I)
					& = \mathit{wp}(\mathit{body}, I).
	\end{array}
	$$
\end{proof}

\section{Experiment Details}\label{subsec:exm}

\subsection{Non-linear Probabilistic Programs}\label{subsec:nonl}
We use a non-linear probabilistic program to model a \emph{simple perceptron,}
which is an algorithm for supervised learning of binary classifiers in machine learning.
It gives a linear classifier function to decide
whether an input belongs to one class or another
based on a set of given data.
Assume the training data is the collection of pairs $(x_i, y_i)_{i \in I}$, where $y_i$ is the desired output value of $x_i$.
We have to learn the linear function $f(x)$ that maps the data to a single binary value:
$$
f(x_i)=\left\{\begin{array}{cl}
1&\textrm{if } w\cdot x_i+b>0\\
0&\textrm{otherwise}
\end{array}\right.
$$
When the outcome does not match $y_i$,
the random perceptron updates its classifier by
$w \leftarrow w+x_i y_i\ [\eta]\ w$ and $b\leftarrow b+y_i\ [\eta]\ b$
where $\eta$ is a learning rate.
When the input of the perceptron is one data pair $(x,y)$,
the algorithm to generate a simple perceptron can be described as:
\begin{center}
	\fbox{\shortstack[l]{
			real $x$, $y$;\\
			real $w$, $b$;\\
			int $n:=0$;\\
			$\while(y\ (w\cdot x+b)\leq 0)\{$\\
			\ \ \  $w:=w+y\cdot x, b:=b+y\ [\eta]\ \progskip;\ n:=n+1$\\
			\}}}
\end{center}

In our trial, we further set $(x,y)=(1,1)$, $\eta=0.25$ and initialize $w=0$ and $b<0$.
The expected time before the function can correctly classify the input is $E(n)=-2b$, as the method shows.

\subsection{Probabilistic Program with Real Variables}

The figures in this example are based on aviation statistics in 2015,
collected by the Civil Aviation Administration of China~\cite{Airplane}.
68.3\,\% of the scheduled flights are actually flown.
An airplane takes 2\,h 15\,min from Beijing to Shanghai.
The average delay is 21\,min and can be approximated by a normal distribution.
Assume an airliner is scheduled for this flight $x$ times,
then the total flight time (in minutes) can be calculated by the program:
\begin{center}
	\fbox{\shortstack[l]{
			$h:=0$;\\
			$n:=0$;\\
			$\while(n\leq x)\{$ \\
			\ \ \ $h:=h+135+\mathrm{NormDist}(21,\sigma)\ [0.683]\ \progskip;n:=n+1$\\
			\}}}
\end{center}
where $\mathrm{NormDist}(\mu,\sigma)$ is a normal distribution with average of $\mu$ and standard deviation of $\sigma$.
(For expectations, the value of $\sigma$ is of no importance.)
Since the sum of normal distributions is also a normal distribution, we can calculate that $E(h)=106.548x$,
which can be proved by our prototype, with the synthesized invariant loop $106.548x-106.548n+h$.

Further, we consider a slightly more involved version.
A direct flight from Beijing to Hongkong takes 220\,min with an average delay of 40\,min.
An alternative two-leg route starts from Beijing, stops in Shanghai, and ends in Hongkong.
The first leg takes 135\,min, with an average delay of 21\,min.
The second leg takes another 135\,min, with an average delay of 40\,min.
A passenger takes $x$ flights from Beijing to Hongkong;
if the direct flight is cancelled, s/he stops in Shanghai.
(We assume that at most one of the two routes is cancelled for simplicity.)
The total travel time can be calculated by the program:
\begin{center}
	\fbox{\shortstack[l]{
			$h:= 0$;\\ $n:= 0$;\\
			$\while(n\leq x)\{$\\
			\ \ \  $h:=h+220+\mathrm{NormDist}(40,\sigma)\ [0.683]\ h:=h+270+ \mbox{}$\\
			\hspace{36pt}$\mathrm{NormDist}(21,\sigma)+\mathrm{NormDist}(40,\sigma)$;\\
			\ \ \  $n:=n+1$\\
			\}}}
\end{center}

We can see $E(h)=282.507x$, which can be verified by our method, with the synthesized loop invariant $282.507(x-n)+h$.

\subsection{Parametric Example}\label{sec:pe}
In this section we consider some parametric examples such that we can
observe how our approach scales with the number of variables and the
degree of the templates.

\subsubsection{Parametric Linear Program.}
We first consider a linear program with parameter $n \in \mathbb{N}$.
The program scheme is
\begin{center}
	\fbox{\shortstack[l]{
			$h:= 0$;\\
			$\while(t> 0)\{$\\
			\ \ \ $h:=h+x_1+\cdots +x_n\ [0.5]\ h:=h+x_1+\cdots +x_n +\textrm{UnifDist}(0,2n)$\\
			\ \ \ $t:=t-1$\\
			\}}}
\end{center}
The post-expectation of the program is $h$ and the related pre-expectation is $(\frac{n}{2}+x_1+\cdots+x_n)t$.

If the degree of the invariant template is chosen to be $2$,
one gets as the synthesized invariant of this parameterized program $h+(\frac{n}{2}+x_1+\cdots+x_n)t$.
In Table~\ref{tab:deg2}, we observe that the number of coefficients
remaining to be determined by the SDP tool
is quadratic to the number of variables.
Moreover, curve fitting shows the running time is approximately cubic in the number of variables.
\renewcommand\arraystretch{1.2}
\begin{table}[hbt]
	\caption{\label{tab:deg2}Running time with degree $2$ for the parametric linear program.
	The row ``Number of coefficients'' shows the numbers of coefficients in the invariant template remaining to be determined.
	The row ``solvertime'' describes the time SDP solver takes to solve the constraint solving problem.}
	\centering
	\begin{tabular}{|m{100pt}|c|c|c|c|c|c|}
		\hline
		Number of variables&$\ n=15\ $ &$\ n=20\ $ &$\ n=25\ $ & $\ n=30\ $ & $\ n=35\ $ & $\ n=40\ $ \\\hline
		Number of coefficients&$136$&$231$&$351$&$496$&$741$&946\\\hline
		Solvertime&$0.41$\,s&$1.30$\,s&$2.44$\,s&$\phantom{0}8.30\,s$&$20.56$\,s&$46.62$\,s\\ \hline
		Total time&$1.11$\,s&$2.26$\,s&$4.14$\,s&$          12.58\,s$&$22.53$\,s&$78.40$\,s\\ \hline
	\end{tabular}
\end{table}

\subsubsection{Quadratic Program.}
We consider the quadratic program obtained from the above program by
replacing $x_1$ by $x_1^2$. We need a template of degree $4$.
In Table~\ref{tab:deg4} we observe that the number of coefficients is
cubic in the number of variables.
Runtime grows rapidly when variables are being added.
Additionally, when we look at the case with $n=30$ in Table~\ref{tab:deg2} and $n=8$ in Table~\ref{tab:deg4},
we see that for a similar number of variables the latter solvertime is still 50\,\% higher;
so the increase in running time is not only due to the number of coefficients in the invariant template.
One reason is that the constraint matrix in the SDP problem becomes much coarser,
which makes it more difficult to solve the problem.

\begin{table}[hbt]
	\centering
	\caption{\label{tab:deg4}Running time for the parametric polynomial program with degree $4$}
	\begin{tabular}{|m{100pt}|c|c|c|c|c|c|}
		\hline
		Number of variables&$\ n=5\ $&$\ n=6\ $&$\ n=7\ $&$\ n=8\ $&$\ n=9\ $&$\ n=10\ $\\\hline
		Number of coefficients&126&210&330&495&715&1001\\\hline
		Solvertime&$0.96$\,s&$1.59$\,s&$4.29$\,s&$12.66$\,s&$29.42$\,s&$\phantom{0}96.24$\,s\\ \hline
		Total time&$1.43$\,s&$4.57$\,s&$5.27$\,s&$14.98$\,s&$36.04$\,s&$          107.30$\,s\\ \hline
	\end{tabular}
\end{table}

\subsubsection{Polynomial Program.}
Our next trial is to consider the following parameterized version of Example
\ref{exa:rm}.

\[
\langle xy^n-x^2\rangle\ z:=0;\ \while(0<x<y^n)\{x:=x+1\ [0.5]\ x:=x-1;z:=z+1;\}\ \langle x\rangle
\]
with $n \in \mathbb{N}$ as a parameter. The relevant invariant template has degree $2n$.
Table~\ref{tab:ruinn} shows the running times dependent on $n$.

\begin{table}[hbt]
	\centering
	\caption{\label{tab:ruinn}Running time for parameterized version of Example~\ref{exa:rm}}
	\begin{tabular}{|m{100pt}|c|c|c|c|c|c|}
		\hline
		Parameter&$\ n=6\ $&$\ n=7\ $&$\ n=8\ $&$\ n=9\ $&$\ n=10\ $&$\ n=11\ $\\\hline
		Number of coefficients&455&680&969&1330&1771&2300\\\hline
		Solvertime&$19.83$\,s&$64.63$\,s&$182.69$\,s&$470.43$\,s&$1099.2$\,s&$2223.5$\,s\\ \hline
		Total time&$23.19$\,s&$75.02$\,s&$213.90$\,s&$498.52$\,s&$1153.1$\,s&$2431.6$\,s\\ \hline
	\end{tabular}
\end{table}

The number of coefficients grows almost cubic in $n$, 
but the running time grows much faster than in Table~\ref{tab:deg4}, 
which indicates that degree is the major influence on running time of our method.

As a conclusion, the main influence on the running time of our method is the degree of the invariant template.
The number of coefficients also has an influence.

\subsection{Nested Loop program}
We copy an example from~\cite{ChatterjeePOPL16} for analysis of almost-sure termination.
Here we try to generate a loop invariant for the program.

\begin{center}
	\fbox{\shortstack[l]{
			real $x$, $y$;\\
			int $k=0$;\\
			\while($x\leq m$)\{\\
			\ \ \  $y=0$;\\
			\ \ \  \while($y\leq n$)\{\\
			\ \ \  \ \ \  $y=y+\textrm{UnifDist}(-0.1,0.2)$\\	
			\ \ \  \}\\
			\ \ \  $x=x+\textrm{UnifDist}(-0.1,0.2)$;\\
			\ \ \  $k=k+1$\\
			\}
		}}
	\end{center}
	
	Let $\mathit{postE}=k$, for $\mathit{preE}=20(m-x)$, we can synthesize an invariant $I=k+20(m-x)$ as the method shows.

\end{document}